\documentclass[aps,prl,twocolumn,superscriptaddress,showpacs,amsmath,amssymb]{revtex4-1}
\usepackage{amsmath}
\usepackage{amsfonts}
\usepackage{amssymb}
\usepackage{graphicx}
\usepackage{color}

\usepackage{bm}
\usepackage{braket}
\usepackage{gensymb}

\usepackage{hyperref}

\begin{document}

\title{Emergent $\mathrm{SU}(4)$ Symmetry in $\alpha$-ZrCl$_3$ and Crystalline Spin-Orbital Liquids}

\author{Masahiko G. Yamada}
\email[]{m.yamada@issp.u-tokyo.ac.jp}
\affiliation{Institute for Solid State Physics, University of Tokyo, Kashiwa 277-8581, Japan.}
\author{Masaki Oshikawa}
\affiliation{Institute for Solid State Physics, University of Tokyo, Kashiwa 277-8581, Japan.}
\author{George Jackeli}
\altaffiliation[]{Also at Andronikashvili Institute of Physics, 0177
Tbilisi, Georgia.}
\affiliation{Institute for Functional Matter and Quantum Technologies, 
University of Stuttgart, Pfaffenwaldring 57, D-70569 Stuttgart, Germany.}
\affiliation{Max Planck Institute for Solid State Research, Heisenbergstrasse 1, D-70569 Stuttgart, Germany.}
\date{\today}

\begin{abstract}
While the enhancement of the spin-space symmetry from 
the usual $\mathrm{SU}(2)$ to $\mathrm{SU}(N)$ is promising
for finding nontrivial quantum spin liquids,
its realization in magnetic materials remains challenging.
Here we propose a new mechanism by which the $\mathrm{SU}(4)$ symmetry emerges
in the strong spin-orbit coupling limit.
In $d^1$ transition metal compounds with edge-sharing anion octahedra,
the spin-orbit coupling gives rise to strongly bond-dependent and
apparently $\mathrm{SU}(4)$-breaking hopping
between the $J_\textrm{eff}=3/2$ quartets.
However, in the honeycomb structure, a gauge transformation
maps the system to an $\mathrm{SU}(4)$-symmetric Hubbard model.  In the strong repulsion limit
at quarter filling, as realized in $\alpha$-ZrCl$_3,$
the low-energy effective model is the $\mathrm{SU}(4)$ Heisenberg model
on the honeycomb lattice, which cannot have a trivial
gapped ground state and is expected to host a gapless spin-orbital liquid.
By generalizing this model to other three-dimensional lattices,
we also propose crystalline spin-orbital liquids
protected by this emergent $\mathrm{SU}(4)$ symmetry and space group symmetries.
\\ \\
PhySH: Frustrated magnetism, Spin liquid, Quantum spin liquid
\end{abstract}

\maketitle

\textit{Introduction}. ---
Nontrivial quantum spin liquids (QSLs) are expected to exhibit many
exotic properties such as fractionalized
excitations~\cite{Balents2010,Savary2017},
in addition to the absence of the long-range order.
Despite the vigorous studies in the last several decades, however,
material candidates for such QSLs are still rather limited.

An intriguing scenario to realize a nontrivial QSL is by
generalizing the spin system, which usually
consists of spins representing the $\mathrm{SU}(2)$ symmetry, to $\mathrm{SU}(N)$ ``spin'' systems
with $N > 2$.
We expect stronger quantum fluctuations in $\mathrm{SU}(N)$ spin systems
with a larger $N$, which could lead the system to an $\mathrm{SU}(N)$ QSL
even on unfrustrated, bipartite lattices, including the honeycomb
lattice~\cite{Li1998,Hermele2011,Corboz2012,Lajko2013}.

The $\mathrm{SU}(N)$ spin systems with $N > 2$
can be realized in ultracold atomic systems,
using the nuclear spin degrees of freedom~\cite{Cazalilla2014}.
In electron spin systems, however, realization of this $\mathrm{SU}(N)$ symmetry
is more challenging.
It would be possible to combine the spin and orbital
degrees of freedom, so that local electronic states are identified
with a representation of $\mathrm{SU}(N)$.
QSL realized in this context
may be called quantum spin-orbital liquids
(QSOLs) because it involves spin and orbital degrees of freedom.
Despite the appeal of such a possibility,
the actual Hamiltonian is usually not $\mathrm{SU}(N)$-symmetric,
reflecting the different physical origins of the spin
and orbital degrees of freedom.
For example, the relevance of an $\mathrm{SU}(4)$ QSOL has been discussed
for Ba$_3$CuSb$_2$O$_9$ (BCSO) with a decorated honeycomb
lattice structure~\cite{Zhou2011,Nakatsuji2012,Corboz2012}.
It turned out, however, that the estimated parameters for BCSO are
rather far from the model
with an exact $\mathrm{SU}(4)$ symmetry~\cite{Smerald2014}.
Moreover, the spin-orbit coupling (SOC) and the directional dependence of the orbital hopping usually
break both the spin-space and orbital-space $\mathrm{SU}(2)$ symmetries,
as exemplified in iridates~\cite{Jackeli2009}.
Thus, it would seem even more
difficult to realize an $\mathrm{SU}(N)$-symmetric system in real magnets with SOC.
(See Refs.~\cite{Ohkawa1983,Shiina1997,Wang2009,Kugel2015} for proposed
realization of $\mathrm{SU}(N)$ symmetry. However, they do not lead
to QSOL because of their crystal structures.)

In this Letter, we demonstrate a novel mechanism for realizing an $\mathrm{SU}(4)$
spin system in a solid-state system with an onsite SOC.  Paradoxically, the
symmetry of the spin-orbital space can be \emph{enhanced} to $\mathrm{SU}(4)$
when the SOC is strong.
In particular, we propose
$\alpha$-ZrCl$_3$~\cite{Swaroop1964Chem,Swaroop1964Phys,Brauer1978} as the first candidate for an
$\mathrm{SU}(4)$-symmetric QSOL on the honeycomb lattice.  Its $d^1$ electronic
configuration in the octahedral ligand field, combined with the strong
SOC, implies that the ground state of the electron is described by a
$J_\textrm{eff}=3/2$ quartet~\cite{Romhanyi2017}.  In fact, the
resulting effective Hamiltonian appears to be anisotropic in the quartet
space.  Nevertheless, we show that the model is gauge-equivalent to an
$\mathrm{SU}(4)$-symmetric Hubbard model.  In the strong repulsion limit, its
low-energy effective Hamiltonian is the Kugel-Khomskii
model~\cite{Kugel1982} on the honeycomb lattice, exactly at the $\mathrm{SU}(4)$
symmetric point:
\begin{equation}
        H_\textrm{eff} = J \sum_{\langle ij \rangle} \Bigl(\bm{S}_i
\cdot \bm{S}_j+\frac{1}{4}\Bigr)\Bigl(\bm{T}_i \cdot
\bm{T}_j+\frac{1}{4}\Bigr), \label{Eq.KK_SU4}
\end{equation}
where $J>0,$ and $\bm{S}_j$ and $\bm{T}_j$ are pseudospin-$1/2$
operators defined for each site $j$.
The $\mathrm{SU}(4)$ symmetry can be made manifest by rewriting the Hamiltonian,
up to a constant shift, as
$ H_\textrm{eff} = \frac{J}{4} \sum_{\langle ij \rangle} P_{ij}$,
where the spin state at each site forms the fundamental representation
of $\mathrm{SU}(4),$ and $P_{ij}$ is the operator which swaps the states at sites $i$
and $j$.  This is a natural generalization of the antiferromagnetic $\mathrm{SU}(2)$ Heisenberg model to $\mathrm{SU}(4).$

The ground state of the $\mathrm{SU}(2)$ spin-1/2 antiferromagnet on the
honeycomb lattice is simply N\'{e}el-ordered~\cite{Roger1989,Fouet2001},
reflecting the unfrustrated nature of the lattice.  On the other hand,
the $\mathrm{SU}(N)$ generalization of the N\'{e}el state by putting different
flavors on neighboring sites gives a macroscopic number of classical
ground states when $N>2$~\cite{Hermele2009,Gorshkov2010,Lajko2017},
implying its instability.
In fact, it was argued that
the $\mathrm{SU}(4)$ antiferromagnet on the honeycomb lattice has a QSOL ground
state without any long-range order~\cite{Corboz2012,Lajko2013}.

\begin{figure}
\centering
\includegraphics[width=8.6cm]{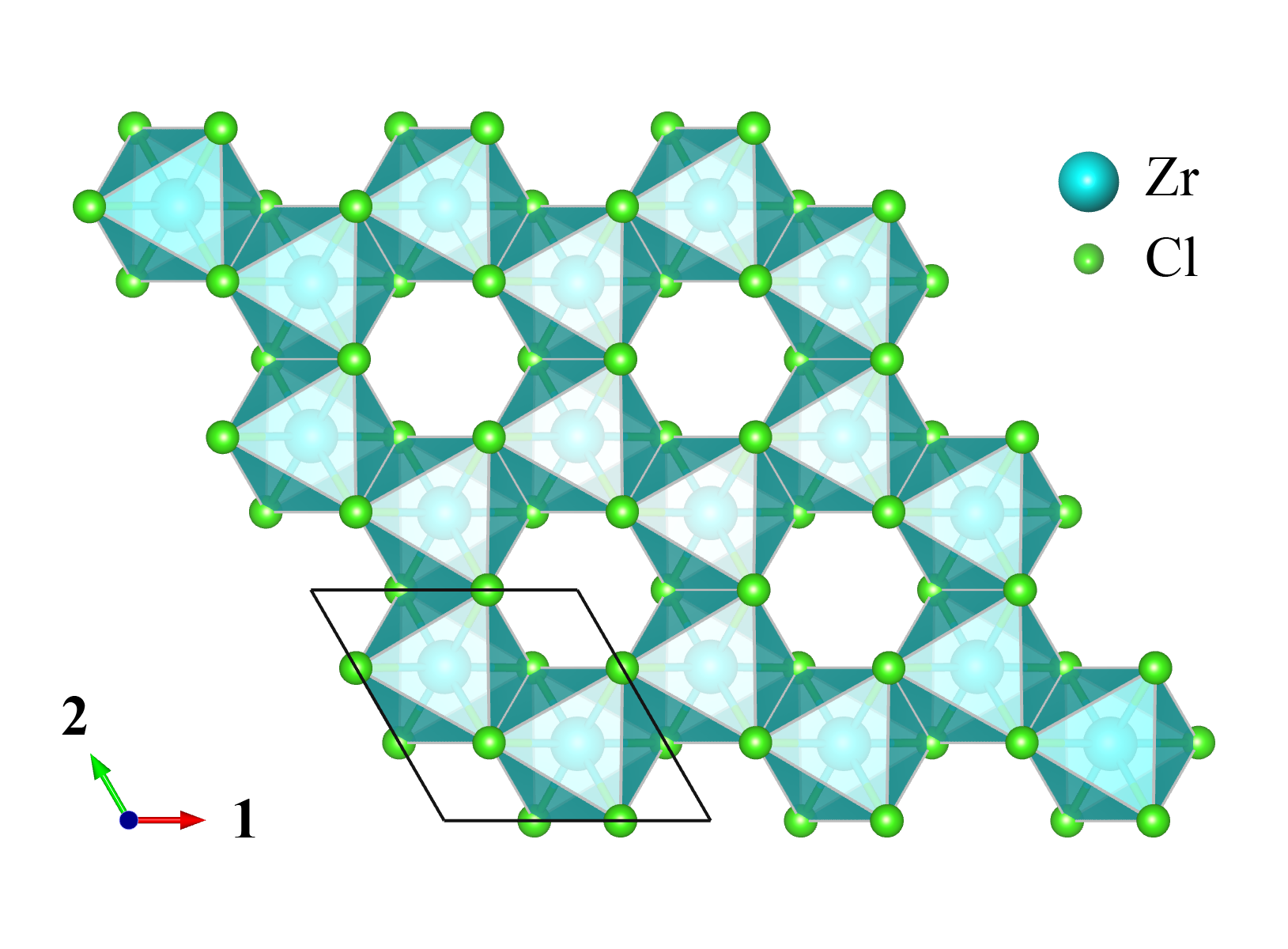}
\caption{Geometric structure of honeycomb $\alpha$-ZrCl$_3.$
Cyan and light green spheres represent Zr and Cl, respectively.
The crystallographic axes are shown and labelled as the 1- and 2-directions.
}
\label{zrcl}
\end{figure}

\textit{Candidate materials}. ---
As we mentioned in the Introduction, we propose
$\alpha$-ZrCl$_3$ with a honeycomb geometry as the first candidate for
the $d^1$ honeycomb system, as shown in Fig.~\ref{zrcl}.
More generally, we consider the class of materials
$\alpha$-$MX_3$, with $M=$ Ti, Zr, Hf, etc., $X=$ F, Cl, Br, etc.
Their crystal structure is almost the same
as that of $\alpha$-RuCl$_3$,
which is known to be an approximate realization of the Kitaev
honeycomb model~\cite{Kitaev2006,Plumb2014}.
However, the electronic structure of $\alpha$-$MX_3$
is different from $\alpha$-RuCl$_3$: 
here, $M$ is in the $3+$ state with a $d^1$ electronic configuration in the
octahedral ligand field.
Our strategy for the realization of $\mathrm{SU}(4)$ spin models starts with a low-energy
quartet of electronic states with the
effective angular momentum $J_\textrm{eff}=3/2$ on each $M$.

For this description to be valid, the SOC has to be strong enough.  As
the atomic number increases from Ti to Hf, SOC gets stronger and the
description by the effective angular momentum becomes exact.
The compounds $\alpha$-$M$Cl$_3$ with $M=$ Ti, Zr and related Na$_2$VO$_3$
have been already reported experimentally.
For $\alpha$-TiCl$_3,$ a structural transition and opening
of the spin gap at $T=217$ K have been reported~\cite{Ogawa1960}.
This implies a small SOC, as it is consistent with
a massively degenerate manifold of spin-singlets
expected in the limit of a vanishing SOC~\cite{Jackeli2007}. 
In compounds with heavier elements, the strong SOC can convert
this extensively degenerate manifold of product states into
a resonating quantum state. 
Thus, we expect realization of the $\mathrm{SU}(4)$ QSOL due to strong
SOC with metal ions heavier than Ti.  In the following, we pick up
$\alpha$-ZrCl$_3$ as an example, although the same analysis should apply
to $\alpha$-HfCl$_3,$ and $A_2M^\prime$O$_3$ ($A=$ Na, Li, etc.,
$M^\prime=$ Nb, Ta, etc.) as well.

\begin{figure}
\centering
\includegraphics[width=8.6cm]{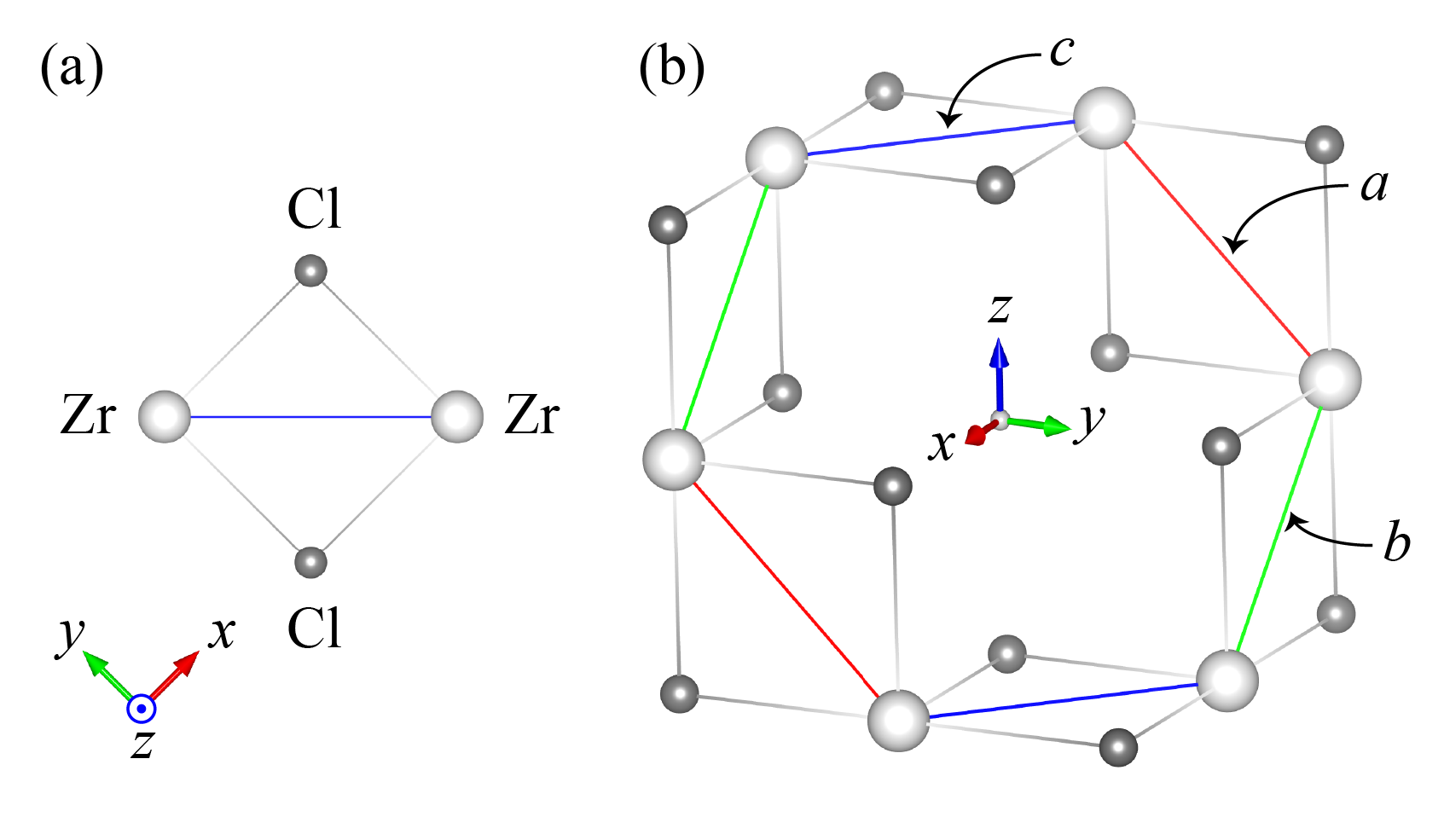}
\caption{(a) Superexchange pathways between two Zr ions connected by
a $c$-bond (blue) in $\alpha$-ZrCl$_3.$
White and grey spheres represent Zr and Cl atoms, respectively.
(b) Three different types of bonds in $\alpha$-ZrCl$_3.$
Red, light green, and blue bonds represent $a$-, $b$-, and $c$-bonds on the
$yz$-, $zx$-, and $xy$-planes, respectively.}
\label{honeycomb}
\end{figure}

\textit{Effective Hamiltonian}. ---
In the strong ligand field, the description with one electron
in the threefold degenerate $t_{2g}$-shell for $\alpha$-ZrCl$_3$ becomes exact.
We denote these $d_{yz}$, $d_{zx}$, and $d_{xy}$-orbitals
by $a,$ $b,$ $c,$ respectively.
Let $a_{j\sigma},$ $b_{j\sigma}$ and $c_{j\sigma}$ represent annihilation operators on these orbitals on the $j$-th site of Zr$^{3+}$ with spin-$\sigma$, and
$n_{\xi\sigma j}$ with $\xi \in \{a,b,c\}$ be the corresponding number operators.
We also use this $(a,\,b,\,c)=(yz,\,zx,\,xy)$ notation to label bonds:
each Zr --- Zr bond is called $\xi$-bond ($\xi=a,$ $b,$ $c$)
when the superexchange pathway is on the $\xi$-plane~\footnote{The Cartesian $xyz$ axes are defined as
in Fig.~\ref{honeycomb}(b).}, as illustrated in Fig.~\ref{honeycomb}.

We define a $J_\textrm{eff}=3/2$ quartet spinor as $\psi = (\psi_{\uparrow \uparrow},\psi_{\uparrow \downarrow},\psi_{\downarrow \uparrow},\psi_{\downarrow \downarrow})^t = (\psi_{3/2},\psi_{-3/2},\psi_{1/2},\psi_{-1/2})^t ,$ where $\psi_{J^z}$ is the annihilation operator for the $\ket{J=3/2, J^z}$ state.
Assuming the SOC is the largest electronic energy scale, except for the ligand field splitting,
fermionic operators can be rewritten by the quartet
$\psi_{j\tau\sigma}$ as follows.
\begin{align}
        a_{j\sigma}^\dagger &= \frac{\sigma}{\sqrt{6}} (\psi_{j\uparrow \bar{\sigma}}^\dagger-\sqrt{3}\psi_{j\downarrow \sigma}^\dagger), \label{Eq.a} \\
        b_{j\sigma}^\dagger &= \frac{i}{\sqrt{6}} (\psi_{j\uparrow \bar{\sigma}}^\dagger+\sqrt{3}\psi_{j\downarrow \sigma}^\dagger), \label{Eq.b} \\
        c_{j\sigma}^\dagger &= \sqrt{\frac{2}{3}}\psi_{j\uparrow\sigma}^\dagger, \label{Eq.c}
\end{align}
where the indices $\tau$ and $\sigma$ of $\psi_{j\tau\sigma}$ label the pseudoorbital
and pseudospin indices, respectively.
We begin from the following Hubbard Hamiltonian for
$\alpha$-ZrCl$_3,$
\begin{align}
        H =& -t \sum_{\sigma, \langle ij \rangle \in \alpha} (\beta_{i\sigma}^\dagger \gamma_{j\sigma}+\gamma_{i\sigma}^\dagger \beta_{j\sigma})+ h.c. \nonumber \\
        &+ \frac{U}{2} \sum_{j, (\delta,\sigma) \neq (\delta^\prime,\sigma^\prime)} n_{\delta\sigma j}n_{\delta^\prime \sigma^\prime j}, \label{Eq.original}
\end{align}
where $t$ is a real-valued hopping parameter through the hopping shown in Fig.~\ref{honeycomb}(a), $U>0$ is the Hubbard interaction, $\langle ij \rangle \in \alpha$ means that the bond $\langle ij\rangle$ is
an $\alpha$-bond,
$\langle \alpha,\beta,\gamma \rangle$ runs over every cyclic permutation of $\langle a,b,c \rangle,$ and $\delta,\delta^\prime \in \{a,b,c\}.$
By inserting Eqs.~\eqref{Eq.a}-\eqref{Eq.c}, we get
\begin{equation}
        H= -\frac{t}{\sqrt{3}} \sum_{\langle ij \rangle} \psi_i^\dagger U_{ij} \psi_j +h.c.
        + \frac{U}{2} \sum_{j} \psi_j^\dagger \psi_j (\psi_j^\dagger \psi_j-1), \label{Eq.Hub}
\end{equation}
where $\psi_j$ is the $J_\textrm{eff}=3/2$ spinor on the $j$th site,
and $U_{ij}=U_{ji}$ is a $4\times 4$ matrix
\begin{equation}
        U_{ij} = \begin{cases}
    U^a = \tau^y \otimes I_2 & (\langle ij \rangle \in a) \\
    U^b = -\tau^x \otimes \sigma^z & (\langle ij \rangle \in b) \\
    U^c = -\tau^x \otimes \sigma^y & (\langle ij \rangle \in c)
  \end{cases},
\end{equation}
where $I_m$ is the $m \times m$ identity matrix, while
$\bm{\tau}$ and $\bm{\sigma}$ are Pauli matrices
acting on the $\tau$ and $\sigma$ indices
of $\psi_{j\tau\sigma},$ respectively.
We note that $U^{a,b,c}$ are unitary and Hermitian, and thus $U_{ji}={U_{ij}}^\dagger
= U_{ij}$.

Now we consider a (local) $\mathrm{SU}(4)$ gauge transformation,
\begin{equation}
        \psi_j \to g_j\cdot \psi_j, \qquad
        U_{ij} \to g_i U_{ij} g_j^\dagger,
\end{equation}
where $g_j$ is an element of $\mathrm{SU}(4)$ defined for each site $j$.
For every loop $C$ on the lattice, the $\mathrm{SU}(4)$ flux defined by the product $\prod_{\langle ij \rangle \in C} U_{ij}$ is invariant under the gauge transformation.

Remarkably, 
for each elementary hexagonal loop (which we call plaquette) $p$
in the honeycomb lattice with the coloring illustrated
in Fig.~\ref{honeycomb}(b),
\begin{equation}
        \prod_{\langle ij \rangle \in p} U_{ij}=U^a U^b U^c U^a U^b U^c=(U^a U^b U^c)^2 =-I_4,
\end{equation}
which corresponds to just an Abelian phase $\pi$.
Since all the flux operators on the honeycomb lattice can be made of some product of
these plaquettes,
there is an $\mathrm{SU}(4)$ gauge transformation to reduce
the model~\eqref{Eq.Hub} to the $\pi$-flux
Hubbard model $H$ with a global $\mathrm{SU}(4)$ symmetry, as proven in Sec.~A of SM~\cite{SM}.
\begin{equation}
        H = -\frac{t}{\sqrt{3}} \sum_{\langle ij \rangle} \eta_{ij} \psi_i^{\dagger} \psi_j + h.c.
        + \frac{U}{2} \sum_{j} \psi_j^{\dagger} \psi_j (\psi_j^{\dagger} \psi_j -1), \label{Eq.piflux}
\end{equation}
where the definition of $\eta_{ij}=\pm 1$, arranged
to insert a $\pi$ flux inside each plaquette, is included in Sec.~A of SM~\cite{SM}.
At quarter filling, i.e. one electron per site,
which is the case in $\alpha$-ZrCl$_3,$ the system
becomes a Mott insulator for a sufficiently large $U/|t|$.
The low-energy effective Hamiltonian for the spin and orbital degrees of
freedom, obtained by the second-order perturbation theory in $t/U,$
is the Kugel-Khomskii model exactly at the $\mathrm{SU}(4)$ point~\eqref{Eq.KK_SU4},
with $\bm{S}=\bm{\sigma}/2,$ $\bm{T}=\bm{\tau}/2,$ and $J=8t^2/(3U)$
in the transformed basis set.
We note that the effective Hamiltonian does not depend on the phase
factor $\eta_{ij}$, as it cancels out in the second-order perturbation
in $t/U$.
Corboz \textit{et al.} argued that this $\mathrm{SU}(4)$ 
Heisenberg model on the honeycomb lattice hosts a gapless
QSOL~\cite{Corboz2012}.
Therefore, we have found a possible realization of gapless QSOL 
in $\alpha$-ZrCl$_3$ with an \textit{emergent} $\mathrm{SU}(4)$ symmetry.

The nontrivial nature of this model
may be understood in terms of
the Lieb-Schultz-Mattis-Affleck (LSMA) theorem for 
the $\mathrm{SU}(N)$ spin systems~\cite{LSM1961,Affleck1986,Lajko2017,YHO2018},
generalized to higher dimensions~\cite{LSM1961,Affleck1988,Oshikawa2000,Hastings2005,Totsuka2017}.
As a result,
under the $\mathrm{SU}(N)$ symmetry and the translation symmetry,
the ground state of the $\mathrm{SU}(N)$ spin system with $n$ spins
of the fundamental representation  per unit cell cannot be unique,
if there is a non-vanishing excitation gap and $n/N$ is not an integer.
This rules out a featureless Mott insulator phase,
which is defined as a gapped phase with a unique ground state,
namely without any spontaneous symmetry breaking or topological order.

For the honeycomb lattice $(n=2)$ there is no LSMA constraint
for an $\mathrm{SU}(2)$ spin system~\cite{Jian2016}.
Nevertheless, for the $\mathrm{SU}(4)$ spin system
we discuss in this Letter, a two-fold ground-state degeneracy
is required to open the gap.
This suggests the stability of a gapless QSOL phase of
the $\mathrm{SU}(4)$ Heisenberg model on the honeycomb lattice.
Especially, assuming the $\pi$-flux Dirac spin-orbital liquid ansatz
proposed in Ref.~\cite{Corboz2012} is correct, a mass gap for the Dirac spectrum
is forbidden unless the $\mathrm{SU}(4)$ or translation symmetry is broken.
Detailed analysis based on the LSMA theorem will be discussed
in a separate publication~\cite{FP}.

\begin{figure}
\centering
\includegraphics[width=8.6cm]{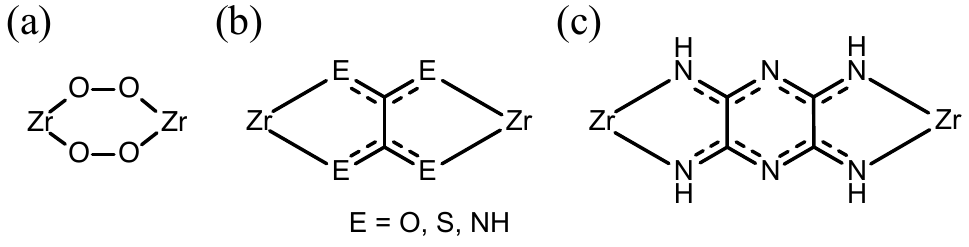}
\caption{Other possible superexchange pathways between two metal ions.
(a) Zr --- O --- O --- Zr.
(b) Oxalate-based metal-organic motif. ($E=$ O, S, NH.)
(c) Tetraaminopyrazine-bridged metal-organic motif.}
\label{super}
\end{figure}

\textit{Other possible structures}. ---
In addition to three-dimensional (3D) inorganic polymorphs~\cite{SM},
metal-organic frameworks (MOFs) with motifs listed in Fig.~\ref{super}
are an interesting playground to explore a variety of $\mathrm{SU}(4)$ QSOLs.
It was recently argued~\cite{Yamada2017MOF} that
Kitaev spin liquids can be realized
in MOFs by a mechanism similar to the one in iridates~\cite{Jackeli2009}.
Since the present derivation of an emergent $\mathrm{SU}(4)$ symmetry shares
the same $t_{2g}$ hopping model as in Ref.~\cite{Jackeli2009}, it is also expected
to apply to Zr- or Hf-based MOFs.
While Fig.~\ref{super}(a) is the longer superexchange pathways expected
in oxides similar to triangular iridates~\cite{Catuneanu2015},
Fig.~\ref{super}(b) and (c) show the superexchange pathways possible in
Zr- or Hf-based MOFs.  With these oxalate- or tetraaminopyrazine-based
ligands, we can expect the two independent superexchange pathways
similar to $\alpha$-ZrCl$_3$ as discussed in
Ref.~\cite{Yamada2017MOF}.

Following the case of the honeycomb lattice, we can repeat the same analysis
to derive the effective spin-orbital model for each 3D tricoordinated lattice.
Recently, the classification of spin liquids on various tricoordinated lattices
attracts much attention, so it is worth investigating~\cite{Hermanns2015Weyl,Hermanns2015BCS,Obrien2016}.
All the tricoordinated lattices considered in this Letter are listed in
Table~\ref{lattice}.  The Table is based on the classification of tricoordinated
nets by Wells~\cite{Wells1977}.  We use a Schl\"afli symbol $(p,c)$ to label a lattice,
where $p$ is the shortest elementary loop length of the lattice,
and $c=3$ means the tricoordination of the vertices.
For example, (6,3) is the two-dimensional (2D) honeycomb lattice, and
all the other
lattices are 3D tricoordinated lattices, distinguished by additional letters
following Wells~\cite{Wells1977}.  $8^2.10$-$a$ is a nonuniform lattice and, thus,
the notation is different from the other lattices.

Generalizing the discussion on the honeycomb lattice, if the $\mathrm{SU}(4)$ flux for
any loop $C$ is reduced to an Abelian phase $\zeta_C$ as
$    \prod_{\langle ij \rangle \in C} U_{ij} =
\zeta_C I_4 \quad (\textrm{for}\,^\forall C)$, 
the Hubbard model acquires the $\mathrm{SU}(4)$ symmetry.
We have examined~\cite{SM,FP} this for each lattice in Table~\ref{lattice},
where a checkmark is put on the $\mathrm{SU}(4)$ column
if the above condition holds.
Moreover, in order to form a stable structure with the present mechanism,
the bonds from each site must form 120 degrees and an octahedral coordination.
This condition is again checked for each lattice, and indicated
in the 120\degree~bond column~\cite{Obrien2016} of Table~\ref{lattice}.
We also put a checkmark on the LSMA column,
when the LSMA theorem implies a ground state degeneracy or
gapless excitations for the $\mathrm{SU}(4)$-symmetric Hubbard model. 
For example, the LSMA constraint applies to
the (8,3)-$b$ lattice, since $n/N=6/4$ is fractional.

\begin{table}
        \centering
        \caption{\label{lattice}Tricoordinated lattices discussed in this Letter.
Space groups are shown in number indices. Nonsymmorphic ones are underlined.
$n$ is the number of sites per unit cell.}
        \begin{ruledtabular}
        \begin{tabular}{p{1.2cm}clp{0.7cm}p{0.4cm}p{0.8cm}p{0.9cm}}
                Wells' notation & Lattice name & $\mathrm{SU}(4)$ & \mbox{120\degree} bond & $n$ & Space group & LSMA \\
                \hline
                (10,3)-$a$ & hyperoctagon & \checkmark\footnotemark[1] & \checkmark & 4 & \underline{\textbf{214}} & \checkmark\footnotemark[2] \\
                (10,3)-$b$ & hyperhoneycomb & \checkmark\footnotemark[1] & \checkmark & 4 & \underline{\textbf{70}} & \checkmark\footnotemark[2] \\
                (10,3)-$d$ & $-$ & \checkmark\footnotemark[1] & $-$ & 8 & \underline{\textbf{52}} & \checkmark\footnotemark[2] \\
                (9,3)-$a$ & hypernonagon & $-$ & $-$ & 12 & \textbf{166} & $-$ \\
                $8^2.10$-$a$ & $-$ & \checkmark & \checkmark & 8 & \underline{\textbf{141}} & $-$ \\
				(8,3)-$b$ & hyperhexagon & \checkmark & \checkmark & 6 & \textbf{166} & \checkmark\footnotemark[3] \\
                $-$ & stripyhoneycomb & \checkmark & \checkmark & 8 & \underline{\textbf{66}} & $-$ \\
                (6,3) & 2D honeycomb & \checkmark & \checkmark & 2 & & \checkmark\footnotemark[4]
        \end{tabular}
\end{ruledtabular}
\footnotetext[1]{The product of hopping matrices along every elementary loop is unity, 
resulting in the $\mathrm{SU}(4)$ Hubbard model with zero flux.}
\footnotetext[2]{Nonsymmorphic symmetries of the lattice are enough to protect a QSOL state, i.e. hosting an XSOL state.}
\footnotetext[3]{Although the model has a $\pi$ flux, with an appropriate gauge choice the unit cell is not enlarged.  Therefore, the LSMA theorem straightforwardly applies to the $\pi$-flux $\mathrm{SU}(4)$ Hubbard model.}
\footnotetext[4]{While the standard LSMA theorem is not effective for
the $\pi$-flux $\mathrm{SU}(4)$ Hubbard model here,
the magnetic translation symmetry works to protect a QSOL state~\cite{LRO2017}.}
\end{table}

\textit{Crystalline spin-orbital liquids}. ---
Finally, we would like to discuss the generalization of the concept of
crystalline spin liquids (XSL)~\cite{Yamada2017XSL} to $\mathrm{SU}(4)$-symmetric
systems.  In the context of gapless Kitaev spin liquids as proposed in
Ref.~\cite{Yamada2017XSL}, a crystalline spin liquid is
defined as a spin liquid state where a gapless point (or a gapped
topological phase) is protected not just by the unbroken time-reversal
or translation symmetry, but by the space group symmetry of the lattice.
In the (10,3) lattices listed in Table~\ref{lattice}, the unit cell
consists of a multiple of 4 sites, and thus the generalized LSMA theorem
seems to allow a featureless insulator if we only consider the translation.

Following Refs.~\cite{PTAV2013,WPVZ2015,PWJZ2017}, however,
we can effectively reduce the size
of the unit cell by dividing the unit cell by the nonsymmorphic
symmetry, and thus the filling constraint becomes tighter with a
nonsymmorphic space group.  Even in the (10,3) lattices, the gapless
QSOL state can be protected by the further extension of the LSMA
theorem~\cite{FP}. We call them crystalline spin-orbital liquids
(XSOLs) in the sense that these exotic phases are protected
in the presence of both the $\mathrm{SU}(4)$
symmetry and (nonsymmorphic) space group symmetries.  We put a checkmark
on the LSMA column of Table~\ref{lattice} if either the standard or
extended LSMA theorem applies.

\textit{Discussions}. ---
We found that, as a consequence of the combination of the octahedral ligand
field and SOC, $\mathrm{SU}(4)$ symmetry \emph{emerges} in $\alpha$-ZrCl$_3$.
In addition to the ZrCl$_3$ (or $A_2M^\prime$O$_3$~\cite{SM}) family we have discussed, 
Zr- or Hf-based MOFs could also realize $\mathrm{SU}(4)$ Heisenberg models
on various tricoordinated lattices.
Especially, 3D (10,3)-$a$~\cite{Coronado2001}, (10,3)-$b$~\cite{Zhang2012hyper},
and $8^2.10$-$a$~\cite{ClementeLeon2013,Yamada2017XSL} lattices, as well as the 2D honeycomb
lattice~\cite{Zhang2014}, were already
realized in some MOFs with an oxalate ligand.
Thus we can expect that microscopic
models defined by Eq.~\eqref{Eq.original} on various tricoordinated lattices will apply
in the same way as the honeycomb $\alpha$-ZrCl$_3$ if we replace the metal ions of
these MOFs with Zr$^{3+},$ Hf$^{3+},$ Nb$^{4+},$ or Ta$^{4+}$~\cite{Yamada2017MOF}.

It would be also interesting to investigate $\mathrm{SU}(4)$
Heisenberg models on nontricoordinated lattices.  Especially, on the
lattice with 1 or 3 sites per unit cell, the LSMA theorem can
exclude the possibility of a simply gapped $\mathbb{Z}_2$ spin liquid
and suggests a $\mathbb{Z}_4$ QSOL or new symmetry-enriched
topological phases instead.

Experimentally, muon spin resonance or nuclear magnetic resonance (NMR)
experiments can rule out the existence of long-range magnetic ordering or spin freezing
in the spin sector. In the orbital sector, a possible experimental signature
to observe the absence of orbital ordering or freezing should be
finite-frequency electron spin resonance (ESR)~\cite{Han2015} or
extended X-ray absorption fine structure~\cite{Nakatsuji2012}.
Especially, finite-frequency ESR can observe the dynamical Jahn-Teller (JT) effect~\cite{Nasu2013,Nasu2015},
where the $g$-factor isotropy directly signals the quantum fluctuation between
different orbitals~\cite{Han2015,Bersuker1975,Bible1970}.
This is applicable to our case because of the shape difference in the $J_\textrm{eff}=3/2$
orbitals~\cite{Romhanyi2017}, and the static JT distortion will result in
the anisotropy in the in-plane $g$-factors~\cite{Iwahara2017}~\footnote{We note that
the trigonal distortion existing \textit{a priori} in real materials only splits the degeneracy
between the out-of-plane and in-plane $g$-factors, and the splitting of the two
in-plane modes clearly indicates an additional (e.g. tetragonal) distortion}.
In addition, the specific heat or thermal transport measurements can distinguish between
the gapped and gapless spectra.
The emergent $\mathrm{SU}(4)$ symmetry would result in changing the
universality class of critical phenomena, or in the accidental coincidence between the
time scales of two different excitations for spins and orbitals observed
by NMR and ESR, respectively.

\textit{Note added}. ---
Following the early version of the present paper on arXiv,
a microscopic derivation of the $\mathrm{SU}(4)$ model on the hyperhoneycomb lattice has been
reported~\cite{Natori2018}.

\begin{acknowledgments}
We thank A.~Banisafar, K.~Collins,
K.~Damle, E.~Demler, V.~Dwivedi, S.~Ebihara, D.~E.~Freedman, Y.~Fuji, B.~I.~Halperin,
M.~Hermanns, H.~Katsura, G.~Khaliullin, D.~I.~Khomskii, R.~Kobayashi, M.~Lajk\'o,
L.~Li, F.~Mila, Y.~Nakagawa, J.~Romhanyi, R.~Sano, K.~Shtengel, A.~Smerald, T.~Soejima,
H.~Takagi, T.~Takayama, T.~Senthil, S.~Tsuneyuki, and, especially, I.~Kimchi,
for helpful comments.
The crystal structure was taken from Materials Project.
M.G.Y. is supported by the Materials Education program for the future leaders in Research, Industry, and Technology (MERIT), and by JSPS.
This work was supported by JSPS KAKENHI Grant Numbers JP15H02113, JP17J05736, and JP18H03686, and by JSPS Strategic International Networks Program No. R2604 ``TopoNet''.  We also acknowledge the support of the Max-Planck-UBC-UTokyo Centre for Quantum Materials.
M.G.Y. acknowledges the Quantum Materials Department at MPI-FKF, Stuttgart for 
kind hospitality during his visits.
\end{acknowledgments}

\bibliography{paper}

\end{document}


\title{Supplemental Material for \\ ``Emergent $\mathrm{SU}(4)$ Symmetry in $\alpha$-ZrCl$_3$ and Crystalline Spin-Orbital Liquids''
}

\author{Masahiko G. Yamada}
\affiliation{Institute for Solid State Physics, University of Tokyo, Kashiwa 277-8581, Japan.}
\author{Masaki Oshikawa}
\affiliation{Institute for Solid State Physics, University of Tokyo, Kashiwa 277-8581, Japan.}
\author{George Jackeli}
\altaffiliation[]{Also at Andronikashvili Institute of Physics, 0177
Tbilisi, Georgia.}
\affiliation{Institute for Functional Matter and Quantum Technologies, 
University of Stuttgart, Pfaffenwaldring 57, D-70569 Stuttgart, Germany.}
\affiliation{Max Planck Institute for Solid State Research, Heisenbergstrasse 1, D-70569 Stuttgart, Germany.}
\maketitle

\onecolumngrid

\appendix

In this Supplemental Material, we have
Section~A: Boundary condition effects on the $\mathrm{SU}(N)$ gauge transformation,
Section~B: Hidden $\mathrm{SO}(4)$ symmetry in the Hund coupling, and
Section~C: Flux sectors for various tricoordinated lattices.

\section{Section~A: Boundary condition effects on the $\mathrm{SU}(N)$ gauge transformation}

First, we begin from the one-dimensional (1D) Hubbard model with an open boundary condition (OBC).
\begin{equation}
		H_\textrm{1DOBC}=-t\sum_{j=1}^{L-1} \psi_j^\dagger U_{j,j+1} \psi_{j+1} + h.c. + \frac{U}{2} \sum_{j=1}^L \psi_j^\dagger \psi_j (\psi_j^\dagger \psi_j-1),
\end{equation}
where $L$ is a system size, $\psi_j$ is a $N$-component spinor, $U_{j,j+1}$ is an $N \times N$
unitary matrix defined on the $j$th site, and $t$ and $U$ are real-valued hopping and Hubbard
terms, respectively.
The (local) gauge transformation is simply given by the following string operator $g_j.$
\begin{align}
		g_j &= \prod_{k=1}^{j-1} U_{k,k+1}, \\
        \psi_j^\prime &= g_j\cdot \psi_j, \\
		U_{j,j+1}^\prime &= g_j U_{j,j+1} g_{j+1}^\dagger = I_N,
\end{align}
where $I_m$ is the $m \times m$ identity matrix.  Thus, 1D Hubbard model with OBC is
a trivial case where we can always make it $\mathrm{SU}(N)$-symmetric.
\begin{equation}
		H_\textrm{1DOBC}=-t\sum_{j=1}^{L-1} \psi_j^{\prime\dagger} \psi_{j+1}^\prime + h.c. + \frac{U}{2} \sum_{j=1}^L \psi_j^{\prime\dagger} \psi_j^\prime (\psi_j^{\prime\dagger} \psi_j^\prime-1),
\end{equation}

Therefore, in 1D electronic systems on a linear chain with
nearest-neighbor hoppings only,
if the $N\times N$ hopping matrices are all unitary,
the tight-binding Hubbard model is trivially gauge-equivalent to
the 1D $\mathrm{SU}(N)$ Hubbard model~\cite{Arovas1995,Pati1998,Itoi2000,Kugel2015}.
Such emergence of the $\mathrm{SU}(N)$ symmetry by the gauge transformation
becomes more nontrivial in higher dimensions because there is a
topological obstruction coming from the lattice geometry
and also a possibility to realize topological ground state
degeneracy, which is impossible in 1D systems~\cite{Chen2011}.

Before going to higher dimensions, it is
instructive to consider the 1D Hubbard model with a periodic boundary condition (PBC).
\begin{equation}
		H_\textrm{1DPBC}=-t\sum_{j=1}^{L} \psi_j^\dagger U_{j,j+1} \psi_{j+1} + h.c. + \frac{U}{2} \sum_{j=1}^L \psi_j^\dagger \psi_j (\psi_j^\dagger \psi_j-1),
\end{equation}
where $\psi_{L+1}$ is identified as $\psi_1.$  Clearly the gauge transformation
does not change the flux inside the loop, so there is a necessary condition
to have a gauge transformation which makes the Hamiltonian $\mathrm{SU}(N)$-symmetric,
\begin{equation}
        \prod_{j=1}^L U_{j,j+1} = \zeta I_N,
\end{equation}
with some $|\zeta|=1.$
This is also a sufficient condition.  If we apply the same gauge transformation
$g_j = \prod_{k=1}^{j-1} U_{k,k+1}$ as the OBC case for $j=1,\dots,L,$
the transformed matrices become
\begin{equation}
		U_{j,j+1}^\prime = \begin{cases}
				\prod_{k=1}^L U_{k,k+1} = \zeta I_N & (j=L) \\
				I_N & (\textrm{otherwise})
\end{cases}.
\end{equation}
Thus, the resulting Hamiltonian is completely $\mathrm{SU}(N)$-symmetric with a factor $\zeta,$
\begin{equation}
		H_\textrm{1DPBC}=-t\Bigl( \sum_{j=1}^{L-1} \psi_j^{\prime\dagger} \psi_{j+1}^\prime + \zeta \psi_L^{\prime\dagger} \psi_{1}^\prime \Bigr) + h.c. + \frac{U}{2} \sum_{j=1}^L \psi_j^{\prime\dagger} \psi_j^\prime (\psi_j^{\prime\dagger} \psi_j^\prime-1).
\end{equation}
It must be noted that $\zeta$ cannot be eliminated by any gauge transformation and
thus it is physical and called (magnetic) flux.

\begin{figure}
\centering
\includegraphics[width=12cm]{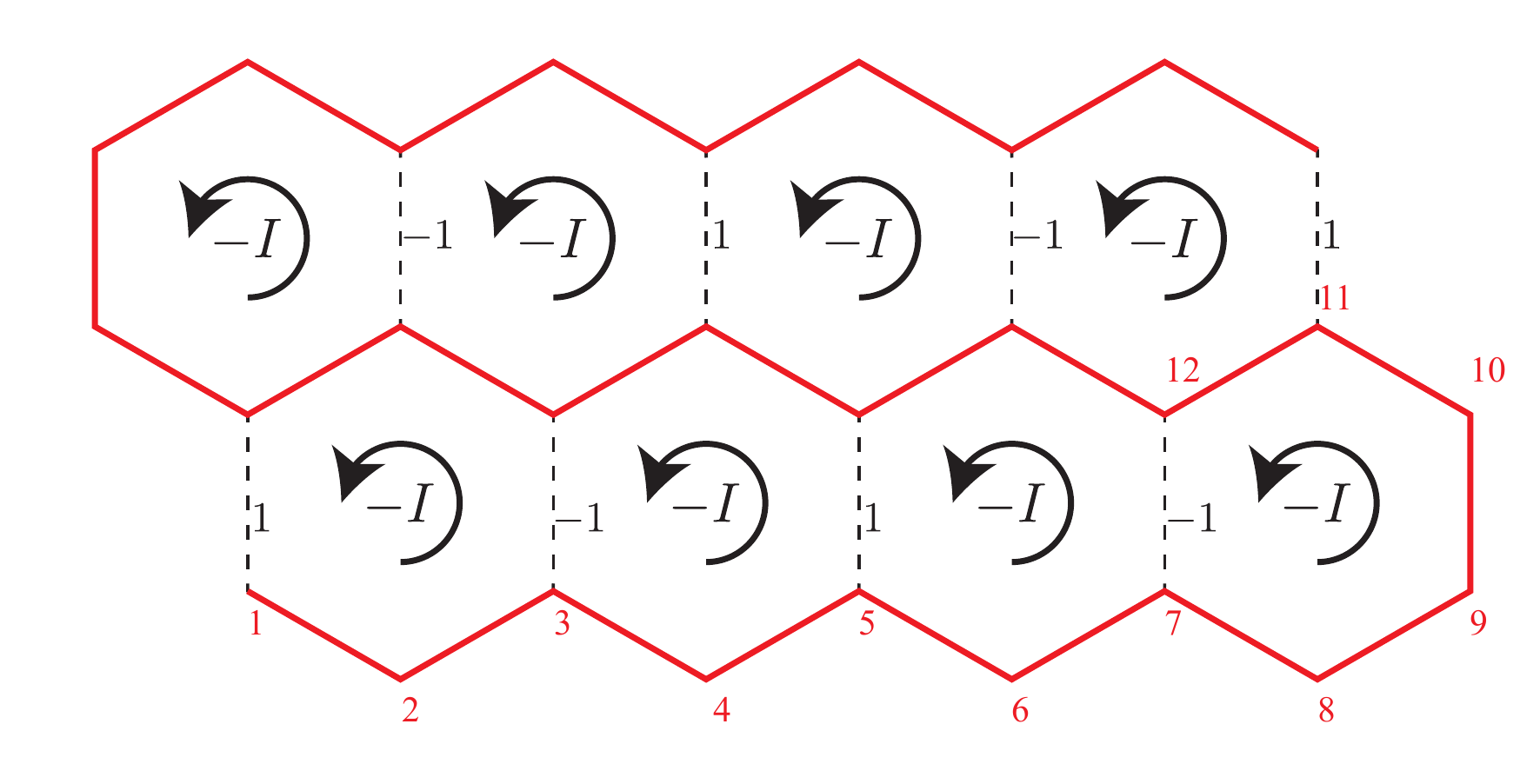}
\caption{Flake of the honeycomb lattice to show how the gauge transformation
works for OBC.  Along the red solid line, we used 1D gauge transformation and
the flux constraints automatically determines the transformed hopping matrices for
the rest of the bonds shown in black dashed lines.}
\label{gauge}
\end{figure}

As for OBC, it is almost trivial to expand the proof of the existence of the
gauge transformation to higher dimensions.  This can be achieved by drawing the lattice with
a single stroke of the brush.  For simplicity, we use the finite-size two-dimensional (2D)
honeycomb lattice
with OBC.  We begin from the following Hamiltonian.
\begin{equation}
		H_\textrm{2D}= -\frac{t}{\sqrt{3}} \sum_{\langle ij \rangle} \psi_i^\dagger U_{ij} \psi_j +h.c.
        + \frac{U}{2} \sum_{j} \psi_j^\dagger \psi_j (\psi_j^\dagger \psi_j-1), \label{Eq.Hub}
\end{equation}
where $U_{ij}$ is again an $N\times N$ unitary matrix defined for each bond,
and $\psi_j$ is the $N$-component spinor on the $j$th site.
Assuming each site is numbered in order
for some nearest-neighbor site to have the subsequent number,
we can do the same gauge transformation as the 1D OBC case.
Again, this gauge transformation does not change the flux value for any loops,
so there is a necessary condition to get a $\mathrm{SU}(N)$-symmetric model for each hexagonal
plaquette (elementary loop) $p.$
\begin{equation}
		\prod_{\langle ij \rangle \in p} U_{ij} = \zeta_p I_N \qquad (\textrm{for}\,^\forall p). \label{Eq.flux}
\end{equation}
This condition is actually sufficient for OBC (assuming the existence of a single stroke
path).  We take a flake of the honeycomb lattice shown in Fig.~\ref{gauge}.
For simplicity, we use $\zeta_p = -1$ for $\alpha$-ZrCl$_3$ as discussed in the main text,
but $\zeta_p$ can generally depend on each plaquette $p.$

If we draw a single stroke path shown as the red solid line in Fig.~\ref{gauge},
all the unitary matrices on the red bonds become identity by the gauge transformation
for the 1D red line.  Remaining are black dashed bonds, but their hopping matrices
are fixed by the flux condition (Eq.~\eqref{Eq.flux}).  In the case of Fig.~\ref{gauge},
around the bottom plaquettes the hopping matrices are determined from right to left
because five of the surrounding matrices are made identity one by one for each plaquette.
By continuing this, all the unitary matrices are transformed into some $\eta_{ij}$ times
identity with $|\eta_{ij}|=1,$ and thus the Hamiltonian becomes completely $\mathrm{SU}(N)$-symmetric. We call this transformed gauge theorists' gauge.
\begin{equation}
		H_\textrm{2D} = -\frac{t}{\sqrt{3}} \sum_{\langle ij \rangle} \eta_{ij} \psi_i^{\prime \dagger} \psi_j^\prime + h.c.
        + \frac{U}{2} \sum_{j} \psi_j^{\prime\dagger} \psi_j^\prime (\psi_j^{\prime\dagger} \psi_j^\prime-1), \label{Eq.piflux}
\end{equation}
where $\eta_{ij}=1$ for red bonds, while the sign of $\eta_{ij}=\pm 1$ depends on each bond
for black dashed bonds as indicated in Fig.~\ref{gauge} by the number near the black dashed bond.
This is nothing but the model called a $\pi$-flux Hubbard model on the honeycomb lattice
and the model can be constructed by changing the sign of the $c$-bonds alternately
along the perpendicular direction.  This gauge transformation effectively doubles
the size of the unit cell.

Finally, we would like to discuss the 2D PBC case.  In this case, we cannot find a
gauge transformation, even if we assume the flux condition (Eq.~\eqref{Eq.flux}) for
every hexagonal plaquette.  The final obstructions
to be considered are global (or topological) ones, which are two types of noncontractible
loops on the 2D torus.  The noncontractible loops in the same homotopy class are related by
the flux conditions, so it is enough to consider only two noncontractible loops $C_1$ and
$C_2$ along the $1$- and $2$-directions, respectively.  Assuming the size of the torus
to be $L_1 \times L_2$ original unit cells, the lengths of $C_1$ and $C_2$ become
multiples of $L_1$ and $L_2,$ respectively.  The necessary and sufficient conditions
to find a gauge transformation in addition to Eq.~\eqref{Eq.flux} are two new
flux conditions for $C_1$ and $C_2,$
\begin{equation}
		\prod_{\langle ij \rangle \in C_1} U_{ij} = \zeta_{C_1} I_N, \qquad \prod_{\langle ij \rangle \in C_2} U_{ij} = \zeta_{C_2} I_N.
\end{equation}

In general these fluxes cannot be Abelian for any sets of unitary matrices $U_{ij}.$
Thus, we specifically consider the model of $\alpha$-ZrCl$_3$ discussed in the main text.
In this model, all the hopping matrices are accidentally written by Pauli matrices, and
their products only take some Pauli matrices times a complex number, which actually
only takes $1,i,-1,-i.$  In other words, their products are included in the Pauli group
on 2 qubits.  In this group, any element to the power of 4 becomes identity, so
the flux inside the two noncontractible loops become trivial if both $L_1$ and $L_2$ are
multiples of 4.  This is a condition to find a gauge transformation to make the model
explicitly $\mathrm{SU}(N)$-symmetric with a symmetric boundary condition, i.e. a boundary
condition where both $C_1$ and $C_2$ have a zero flux.  If we allow a more general
boundary condition with a $\pi$ flux inside $C_1$ or $C_2,$ then the conditions
for $L_1$ or $L_2$ become milder.

Our effective model for the honeycomb $\alpha$-ZrCl$_3$ was derived based on
the superexchange interactions between the Zr$^{3+}$ ions constructed
from its geometry.  However, similar superexchange interactions can also arise
in the other structures listed in Fig.~3 in the main text, or in face-shared systems.
We note that ZrCl$_3$ has some polymorphs and a chain compound $\beta$-ZrCl$_3$
with face-shared Cl octahedra~\cite{Watts1966} can also host
a 1D $\mathrm{SU}(4)$ Heisenberg model~\cite{Kugel2015}.

Since a nonlayered structure of Na$_2$VO$_3$ has already been reported~\cite{Rudorff1956},
we can expect various three-dimensional (3D) polymorphs
of ZrCl$_3$ or $A_2M^\prime$O$_3$ with $A=$ Na, Li and $M^\prime=$ Nb, Ta, similarly
to 3D $\beta$-Li$_2$IrO$_3$~\cite{Takayama2015} and $\gamma$-Li$_2$IrO$_3$~\cite{Modic2014}.

The generalization from the 2D case to the three-dimensional (3D) case is straightforward.
The difference
is that in 3 dimensions not all the fluxes of the plaquettes (or elementary loops in Section~C)
can be determined independently.  This is called volume constraint and will
be discussed in Section~C.

\section{Section~B: Hidden $\mathrm{SO}(4)$ symmetry in the Hund coupling}

It is clear that the first apparent perturbation of an order $J_H/U \sim \mathcal{O}(0.1)$ is
an onsite Hund coupling $J_H.$  There are other possible perturbations like further-neighbor
interactions, but we can expect that such effects are smaller than that of the Hund coupling
similarly to $\alpha$-RuCl$_3.$  Actually, in the Kitaev materials like
$\alpha$-RuCl$_3$ the nearest-neighbor Kitaev interaction and the third-neighbor
Heisenberg interaction are expected to be comparable~\cite{Winter2016},
but this is probably due to fine tuning
happening in the $J_\textrm{eff}=1/2$ manifold and the Kiteav interaction has to be
smaller than the na\"ive superexchange interaction expected in the whole $t_{2g}$ orbitals
because of the destructive interference which cancels out the direct hopping between
the $J_\textrm{eff}=1/2$ manifold~\cite{Jackeli2009}.
In our $J_\textrm{eff}=3/2$ models realized e.g. in $\alpha$-ZrCl$_3,$ such an accidental
reduction of the highest-order contribution
does not occur even in the nearest-neighbor interactions, so we expect the magnetic interaction
in $\alpha$-ZrCl$_3$ is much larger than the dominant Kiteav interaction in $\alpha$-RuCl$_3,$
and thus one- or two-order larger than the third-neighbor Heisenberg interactions in the
case of $\alpha$-ZrCl$_3.$

Next, in order to evaluate the effect of the Hund coupling,
we will change the ordering of the $J_\textrm{eff}=3/2$ bases to compare the model
with a so-called $\mathrm{SO}(5)$-symmetric Hubbard model discussed in the literature
on $S=3/2$ cold atomic systems~\cite{Congjun2003,Congjun2005,Congjun2006},
\begin{equation}
		\psi = (\psi_{3/2},\psi_{1/2},\psi_{-1/2},\psi_{-3/2})^t = (\psi_{\uparrow \uparrow},\psi_{\downarrow \uparrow},\psi_{\downarrow \downarrow},\psi_{\uparrow \downarrow})^t.
\end{equation}
In this basis it is easy to see a hidden $\mathrm{SO}(4)$ symmetry,
which is a subgroup of $\mathrm{SO}(5) \simeq \mathrm{Sp}(4) \subset \mathrm{SU}(4)$
in the original model.

We will now show the Hund coupling in $\alpha$-ZrCl$_3$ actually possesses
the $\mathrm{SO}(5) \simeq \mathrm{Sp}(4)$ symmetry, although the hopping matrices
break a part of this symmetry. If we add a Hund coupling for the hopping model inside
the $t_{2g}$ orbitals~\cite{Georges2013}, the Hamiltonian becomes
\begin{align}
        H =& -t \sum_{\sigma, \langle ij \rangle \in \alpha} (\beta_{i\sigma}^\dagger \gamma_{j\sigma}+\gamma_{i\sigma}^\dagger \beta_{j\sigma})+ h.c. \nonumber \\
		&+ \sum_{j} \left[ \frac{U-3J_H}{2} N_j(N_j-1) -2J_H \bm{s}_j^2 -\frac{J_H}{2} \bm{L}_j^2 +\frac{5}{2}J_H N_j \right],
\end{align}
where $\langle ij \rangle \in \alpha$ means that the bond $\langle ij\rangle$ is
an $\alpha$-bond,
$\langle \alpha,\beta,\gamma \rangle$ runs over every cyclic permutation of $\langle a,b,c \rangle,$ $N_j$ is a number operator, $\bm{s}_j$ is a total spin, and $\bm{L}_j$ is a total
effective angular momentum.
Assuming a strong spin-orbit coupling limit $\lambda \gg |t|,\,J_H,$
we project the Hilbert space onto the $J_\textrm{eff}=3/2$ manifold.
We note that we will ignore doublon/holon excitations with higher energies in the following discussions.
In the original gauge before the gauge transformation,
which we call lab gauge, the projected Hamiltonian becomes
\begin{equation}
        H= -\frac{t}{\sqrt{3}} \sum_{\langle ij \rangle} \psi_i^\dagger V_{ij} \psi_j +h.c.
		+ \sum_{j} \left[ \frac{U-3J_H}{2} \psi_j^\dagger \psi_j (\psi_j^\dagger \psi_j-1)-
		\frac{4}{9}J_H \bm{J}_j^2 +\frac{5}{2}J_H \psi_j^\dagger \psi_j \right],
\end{equation}
where $\bm{J}_j = \bm{s}_j + \bm{L}_j$ is a total effective angular momentum operator
with a condition $J=3/2$ after the projection, and
\begin{equation}
V_{ij} = \begin{cases}
    V^a = \tau^z \otimes \sigma^y = \Gamma^3 & (\langle ij \rangle \in a) \\
    V^b = -\tau^z \otimes \sigma^x = -\Gamma^2 & (\langle ij \rangle \in b) \\
    V^c = -\tau^y \otimes I_2 = \Gamma^1 & (\langle ij \rangle \in c)
  \end{cases}.
\end{equation}
We used $\bm{s}_j=\bm{J}_j/3$ and $\bm{L}_j=2\bm{J}_j/3$ inside the $J_\textrm{eff}=3/2$ manifold
derived from the Wigner-Eckart theorem. Thus, ignoring the hopping terms,
the Hubbard and Hund couplings possess a hidden $\mathrm{SO}(5) \simeq \mathrm{Sp}(4)$ symmetry
in the same way as the $S=3/2$ cold atomic systems with a spin-preserving interaction.

The hopping term partially breaks this $\mathrm{SO}(5)$ symmetry. To see this
we use anticommuting Dirac gamma matrices in Ref.~\cite{Congjun2003} defined as
\begin{equation}
(\Gamma^1,\Gamma^2,\Gamma^3,\Gamma^4,\Gamma^5) = (-\tau^y \otimes I_2, \tau^z \otimes \sigma^x, \tau^z \otimes \sigma^y, \tau^z \otimes \sigma^z, -\tau^x \otimes I_2).
\end{equation}
Gamma matrices $\Gamma^p$ ($p=1,\dots,5$) are forming an $\mathrm{SO}(5)$ vector, which
transforms as a vector in the same rotation for the hidden $\mathrm{SO}(5)$ symmetry
of the Hund coupling. There is no way to eliminate the non-Abelian hopping just by the
$\mathrm{SO}(5) \simeq \mathrm{Sp}(4)$ gauge transformation, but we can rotate
$\mathrm{SO}(5)$ vectors locally to eliminate the bond dependence of the hopping.

For example, we can rotate all $V_{ij}$s to $\Gamma^5$ and then the Hamiltonian
becomes almost uniform up to the same factors $\eta_{ij}=\pm 1$ as discussed in the
previous section:
\begin{equation}
		H = -\frac{t}{\sqrt{3}} \sum_{\langle ij \rangle} \eta_{ij} \psi_i^{\prime \dagger} \Gamma^5 \psi_j^\prime + h.c.
		+ \sum_{j} \left[ \frac{U-3J_H}{2} \psi_j^{\prime \dagger} \psi_j^\prime (\psi_j^{\prime \dagger} \psi_j^\prime-1)-
		\frac{4}{9}J_H \bm{J}_j^{\prime 2} +\frac{5}{2}J_H \psi_j^{\prime \dagger} \psi_j^\prime \right]. \label{eq.so5}
\end{equation}
This model explicitly has a hidden $\mathrm{SO}(4)$ symmetry because $\Gamma^5$ is
invariant under the $\mathrm{SO}(4)$ subgroup of the $\mathrm{SO}(5)$ rotation
which keeps a vector $(0,0,0,0,1)$ invariant.
The last term is constant in the large $(U-3J_H)$ limit at quarter filling,
so the first meaningful contribution of an order $J_H/U \sim \mathcal{O}(0.1)$ would
be the $\mathrm{SO}(4)$-invariant perturbation coming from the term
$(4J_H/9) \bm{J}_j^{\prime 2},$ which separates the degeneracy of the virtual state
with two electrons per site into $J=0$ and $J=2.$ However, this effect is again
$\mathcal{O}(0.1)$ and, thus, we can expect this $\mathrm{SU}(4)$ breaking perturbation
to be negligible.

We note that the $\mathrm{SO}(5) \simeq \mathrm{Sp}(4)$ gauge transformation is just
a subgroup of the $\mathrm{SU}(4)$ gauge transformation, and it is not enough
to go to ``theorists' gauge'' without any non-Abelian hopping matrices.
In fact, Dirac gamma matrices are not included in the generator of the $\mathrm{Sp}(4)$
rotation for $\psi$ and the rotation is generated by $\Gamma^{pq} = -(i/2)[\Gamma^p,\Gamma^q] = -i\Gamma^p \Gamma^q$
($1\leq p,q \leq 5$)~\cite{Congjun2003}. Since the number of gamma matrices is
conserved mod 2 by the $\mathrm{SO}(5) \simeq \mathrm{Sp}(4)$ rotation,
the hopping matrices written by one gamma matrix cannot be rotated to $\mathrm{SO}(5)$
scaler by the $\mathrm{SO}(5)$ gauge transformation, and this is why $\Gamma^5$ cannot
be eliminated in Eq.~\eqref{eq.so5}.

In this analysis, we only considered the extreme limit $\lambda \gg J_H$ for simplicity
to prove that the $\mathrm{SU}(4)$-breaking term comes from the order of
$\mathcal{O}(0.1)$ by employing the $\mathrm{SO}(5)$ gauge transformation intensively.
While we no longer expect the existence of a hidden $\mathrm{SO}(4)$ symmetry in
a general case, it is not difficult to show that in the second-order perturbation the
contribution breaking the original $\mathrm{SU}(4)$ symmetry always involves an
virtual state with an energy higher than the lowest order by $\lambda$ or $J_H.$
Anyway, we can conclude that, as long as we ignore higher order contributions of
$\mathcal{O}(0.1),$ the emergent $\mathrm{SU}(4)$ symmetry would be robust.

\section{Section~C: Flux sectors for various tricoordinated lattices}

\begin{figure}
\centering
\includegraphics[width=10cm]{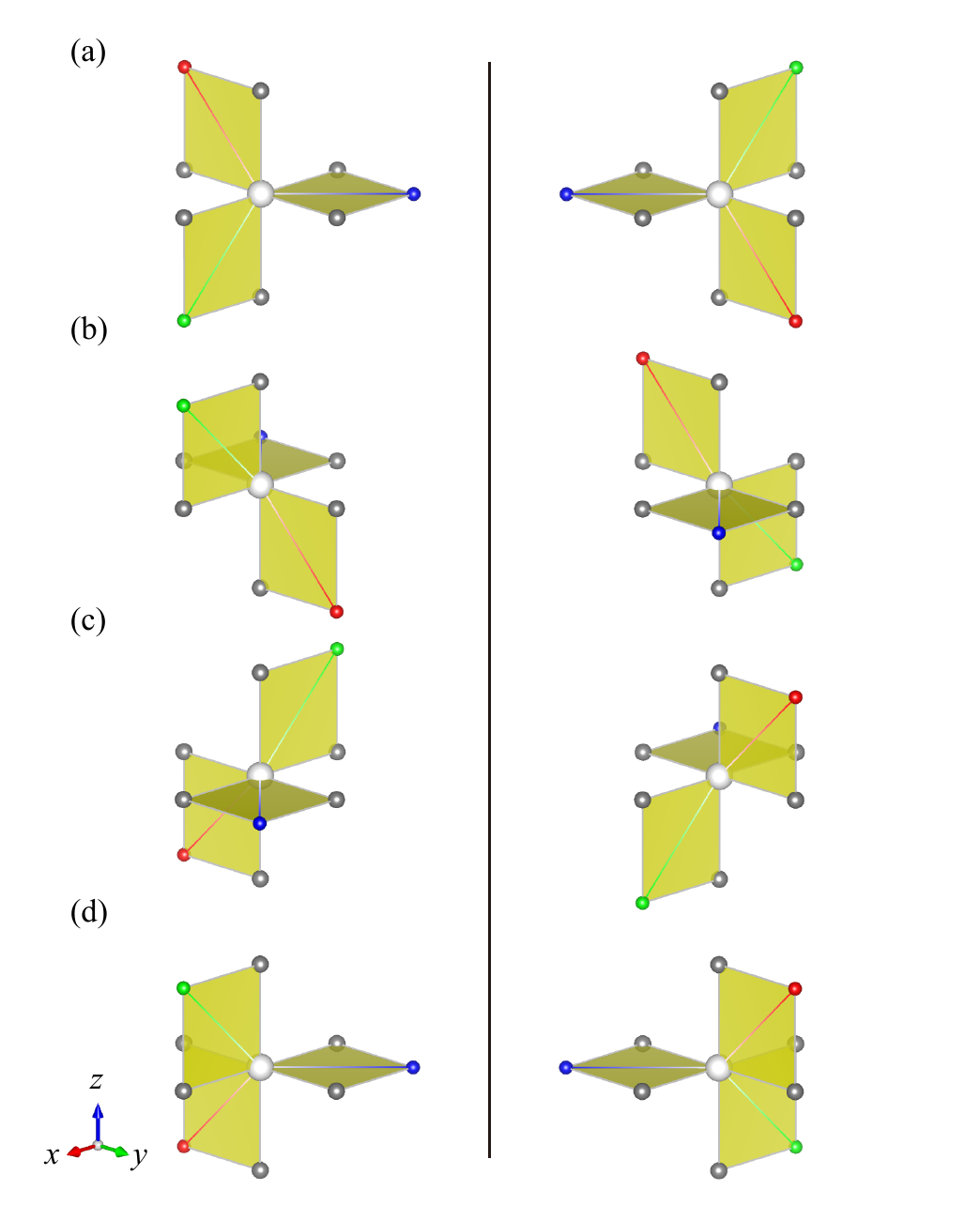}
\caption{All possible ways to connect three bonds in the 3D tricoordinated lattices.
(a) is the same one as that in the 2D honeycomb lattice, while (b), (c), and (d)
are produced by rotating (a) by $180\degree$ around the $x,$ $y,$ and $z$-axes.
The left-hand side and the right-hand side are related by the inversion for
each figure.}
\label{phase}
\end{figure}

The flux sectors for the tricoordinated lattices listed in the main text
can be treated similarly to the Kitaev models on tricoordinated lattices~\cite{Kitaev2006,Obrien2016}
except for the difference in the gauge group.
Following Kitaev~\cite{Kitaev2006}, we use terminology of the lattice gauge theory.
The link variables $U_{ij}$ are Hermitian and unitary (in this case)
$4\times 4$ matrices defined for each bond (link) $\langle ij \rangle$ of the lattice. 
Each link variable depends on its type (color) of the bond as
\begin{equation}
        U_{ij} = \begin{cases}
    U^a = \tau^y \otimes I_2 & (\langle ij \rangle \in a) \\
    U^b = -\tau^x \otimes \sigma^z & (\langle ij \rangle \in b) \\
    U^c = -\tau^x \otimes \sigma^y & (\langle ij \rangle \in c)
  \end{cases}, \label{eq.lab}
\end{equation}
where $\bm{\tau}$ and $\bm{\sigma}$ are independent
Pauli matrices, following the original gauge (basis) used in the main text
(not the one used in the previous section). The bond type $abc$ is determined from
which plane this bond belongs to, as discussed in the main text.
We note that in the 3D case we actually have six types of bonds with additional
$\pm 1$ factors, so $U_{ij} = \pm U^a,\,\pm U^b,\,\pm U^c$ depending on a detailed
structure of the bond $\langle ij \rangle.$ This comes from the spatial dependence of
the sign of the wavefunctions of the $d$-orbitals.

These additional $\pm 1$ factors can simply be gauged out in the following way.
In the 2D honeycomb lattice, there is no sign difference in the same bond type
because all of them are related by the translation symmetry. In some 3D lattices,
even if the two bonds belong to the same type, the hopping matrices can
differ because they are related not by the translation symmetry, but
by the screw or glide symmetry. Accompanied by the reflection or rotation, this symmetry can
actually change the sign of the hopping matrix by $-1$ according to the shape of the
$t_{2g}$-orbitals. When seen from the metal site, it is a $180\degree$ rotation around
the $x,$ $y,$ or $z$-axis. If we consider the signs of the $t_{2g}$-orbitals,
it is clear that $180\degree$ rotation changes the signs of some orbitals,
while the inversion does not change the signs of the $d$-orbitals.
As shown in Fig.~\ref{phase}, there are 8 types of metal sites, and
all of them are related by some $180\degree$ rotation, which causes the sign difference,
up to inversion. Fortunately, however, this additional sign
can be eliminated by some gauge transformation, i.e. local rotations of the definition of the
effective angular momentum $l=1$ of the $t_{2g}$-orbitals.
For example, if the metal site is rotated around the $x$-axis by $180\degree,$
the configuration of the surrounding ligands changes
from Fig.~\ref{phase}(a) to Fig.~\ref{phase}(b). Then, according to the rotation,
we rotate the definition of the angular momentum $l=1$ around the $x$-axis by $180\degree,$
which can be done just by flipping the sign of the $yz$-orbital.
Similarly, for the ones shown in Fig.~\ref{phase}(c), we just flip the
sign of $zx$-orbital. Then, if we connect these two, Fig.~\ref{phase}(b) and (c), along
the $xy$-plane, we obtain an additional $-1$ phase from this gauge transformation, and
it completely cancels out the sign in question.
If we do a similar local rotation in the fictitious orbital space for each metal site
according to the physical $180\degree$ rotation, all the hopping matrices will be returned
to the original ones in Eq.~\eqref{eq.lab}, and after all we do not have to care about
the subtle difference among the same bond type. Thus, Eq.~\eqref{eq.lab} is still valid
after this ``$\mathbb{Z}_2$'' gauge transformation.

\begin{table}
        \centering
		\caption{\label{flux}Flux sector of tricoordinated lattices.  Only the flux value for the shortest elementary loops is shown here.  Nonsymmorphic space group numbers are underlined.  NS means that nonsymmorphic symmetries of the lattice are enough to protect a quantum spin-orbital liquid state.  In addition to the contents of Table~I in the main text, we also include O'Keeffe's three-letter codes~\cite{OKeeffe2003regular,OKeeffe2003semiregular}.}
        \begin{ruledtabular}
                \begin{tabular}{cccccccccc}
                        Wells' & Lattice & O'Keeffe's & Minimal & Flux & 120-degree & Number & \multicolumn{2}{c}{Space group} & LSMA \\
                        notation & name & code & loop length & sector & bond & of sites & symbol & No. & constraints \\
                \hline
				(10,3)-$a$ & hyperoctagon & \textbf{srs} & 10 & 0-flux & \checkmark & 4 & $I4_1 32$ &\underline{\textbf{214}} & NS \checkmark \\
				(10,3)-$b$ & hyperhoneycomb & \textbf{ths} & 10 & 0-flux & \checkmark & 4 & $Fddd$\footnote{The most symmetric case should be $I4_1 /amd,$ including $Fddd$.  Actually, $Fddd$ is enough for the filling constraint.} & \underline{\textbf{70}} & NS \checkmark \\
				(10,3)-$d$ & & \textbf{utp} & 10 & 0-flux & $-$ & 8 & $Pnna$\footnote{There exists another phase with a $Pbcn$ symmetry.  Both symmetries are enough for the filling constraint.} & \underline{\textbf{52}} & NS \checkmark \\
				nonuniform & $8^2.10$-$a$ & \textbf{lig} & 8 & $\pi$-flux & \checkmark & 8 & $I4_1 /amd$ & \underline{\textbf{141}} & $-$ \\
				(8,3)-$b$ & hyperhexagon & \textbf{etb} & 8 & $\pi$-flux & \checkmark & 6 & $R\bar{3}m$ & \textbf{166} & \checkmark \\
				nonuniform & stripyhoneycomb & \textbf{clh} & 6 & $\pi$-flux & \checkmark & 8 & $Cccm$\footnote{There exists a more symmetric phase with a $P4_2/mmc$ symmetry, but it is not enough for the filling constraint.} & \underline{\textbf{66}} & $-$ \\
				(6,3) & 2D honeycomb & \textbf{hcb} & 6 & $\pi$-flux & \checkmark & 2 & & & \checkmark
        \end{tabular}
\end{ruledtabular}
\end{table}

In order to find a gauge transformation
to get an $\mathrm{SU}(4)$ Hubbard model, we have to check that every Wilson loop operator
is Abelian.  In an abuse of language, each Wilson loop will be called flux inside the loop.
We regard a Wilson loop operator $I_4$ as a zero flux, and $-I_4$ as a $\pi$ flux.
In order to get a desired gauge transformation,
it is enough to show the flux inside every elementary loop $C$ is Abelian:
\begin{equation}
        \prod_{\langle ij \rangle \in C} U_{ij} = \zeta_C I_4, \label{Eq.fluxfree}
\end{equation}
with some phase factors $|\zeta_C|=1,$ as discussed in Section~A.

Since $U_{ij}^2=I_4,$ not all the fluxes are independent. In the case of a $\mathbb{Z}_2$ gauge
field, the constraints between multiple fluxes are called volume constraints~\cite{Obrien2016}.
However, due to the non-Abelian nature of the flux structure, it is subtle
whether they apply.  Fortunately, the above $U^\alpha$ ($\alpha=a,\,b,\,c$) obeys
the following anticommutation relations.
\begin{equation}
        \{U^\alpha,U^\beta\} = 2\delta^{\alpha\beta}I_4.
\end{equation}
This algebraic relation proves the product of the fluxes of the loops surrounding some
volume must vanish (volume constraints).  Moreover, we can easily show that, if every
bond color is used even times in each loop, which is a natural consequence for the lattices
admitting materials realization, the flux inside should always be Abelian with
$\zeta_C=\pm 1.$  Actually, every lattice included in Table~\ref{flux} obeys this
condition, so we have already proven all of them have an Abelian flux sector.

The remaining subtle problem is which flux these elementary loops have, a zero flux,
or a $\pi$ flux. To check this, we need to investigate every loop one by one.
To calculate every flux value systematically, we often use space group symmetries to relate
two elementary loops, even though the system is in the strong spin-orbit
coupling limit~\footnote{The threefold rotation symmetry of the $xyz$-axes
of the Cartesian coordinate is not clear in the gauge used in the main text.
The spin quantization
axis along the (111) direction will make this symmetry explicit.}.
We have checked all the elementary loops in the tricoordinated lattices listed
above~\cite{FP}.
Only the flux value for the shortest elementary loops is shown in Table~\ref{flux}.

\bibliography{suppl}